# Combined *in situ* microstructural study of the relationships between local grain boundary structure and passivation on microcrystalline copper

*Mohamed Bettayeb,[a] Vincent Maurice,[a,*] Lorena H. Klein,[a] Linsey Lapeire,[b] Kim Verbeken,[b] and Philippe Marcus[a,*]*

[a] PSL Research University, CNRS - Chimie ParisTech, Institut de Recherche de Chimie Paris (IRCP), Physical Chemistry of Surfaces Group, 11 rue Pierre et Marie Curie, 75005 Paris, France

[b] Department of Materials, Textiles and Chemical Engineering, Ghent University (UGent), Technologiepark 46, 9052 Zwijnaarde (Ghent), Belgium

## Abstract

*In situ* Electro-Chemical Scanning Tunneling Microscopy (ECSTM) and Electron Back-Scatter Diffraction analysis of the same local microstructural region were combined to study the relationships between grain boundary (GB) type and structure and passivation on microcrystalline copper in 0.1 M NaOH aqueous solution. The results show that, for high angle random boundaries, passivation in the Cu(I) oxidation range is characterized by a decrease of the depth of GB edge region due to the formation of a passive film locally thicker than on the adjacent grains and thus to anodic oxidation being locally less efficient since it consumes more copper at the grain boundaries. For Σ3 CSL boundaries, the Cu(I) passivation efficiency was observed to be dependent on the local structure of the boundaries. A transition from more to less efficient passivation was observed for a deviation angle of 0.4-0.5° of the {111} GB plane with respect to the perfect geometry of an ideal coherent twin boundary. This transition would result from the effect on the local anodic oxidation properties of an increased density of misfit dislocations accommodating the increased deviation of the boundary plane from the exact CSL plane.

Keywords: Copper; Electrochemical STM; EBSD; Grain boundary; Structure; Passivation

---

[*] Corresponding authors:
V. Maurice (vincent.maurice@chimie-paristech.fr); P. Marcus (philippe.marcus@chimie-paristech.fr)

# 1. Introduction

Intergranular corrosion is a major form of localized corrosion limiting the durability of polycrystalline metallic materials. In aggressive environment, the corrosive attack of the grain boundary (GB) network is first initiated at the topmost surface and then can penetrate and propagate in the sub-surface, leading to degradation of the microstructure. The resistance to intergranular corrosion can be improved by GB engineering that aims at designing and producing polycrystalline materials, including micro and nanocrystalline metals, with the most resistive GB network. This requires a good knowledge of the relationships between GB type and structure and corrosion resistance that is most often developed by performing micrographic and microscopic characterizations of GB engineered materials submitted to intergranular sub-surface penetrating attack [1-21]. From such studies, coherent twin boundaries are considered as most resistant to intergranular corrosion [3,14]. Coherent twin boundaries are CSL (Coincidence Site Lattice) boundaries with a {111}-oriented GB plane and of Σ3 index, meaning that 1/3 of the lattice sites in the GB plane coincide with the lattice sites of the two grains. Among the low Σ CSL boundaries, only the Σ3 boundaries and partially the Σ9 boundaries have been reported to better resist to intergranular corrosion [6,9,12,14,16]. Higher Σ CSL boundaries and random boundaries, that cannot be described by a coincidence site lattice, are high angle boundaries that do not resist intergranular sub-surface attack [2,6,8,11-14,16].

The relationships between the local structure at the surface termination of the grains boundaries and the resistance in the initial stages of intergranular corrosion, i.e. before penetration and propagation in the sub-surface region, have been much less studied. This is due to the difficulty of applying analytical methods sensitive to the electrochemically-induced local alterations of the topmost surface. Recently, ECSTM (Electro-Chemical Scanning Tunneling Microscopy) has been applied to microcrystalline copper to investigate *in situ* the local corrosion resistance at the surface termination of grain boundaries in metals [22-26]. It was found that dissolution in the active state, i.e. in the absence of formation of a passive film, did not occur at the GB edges assigned to coherent twin boundaries whereas it did at GB edges assigned to other CSL boundaries or to random boundaries in the same conditions of early corrosive attack [22]. Very recently, a deeper insight into the GB behavior in the very first stages of active dissolution was brought using an innovative approach combining analysis of the same local microstructural region by ECSTM and EBSD (Electron Back-Scatter Diffraction) [26]. Random high angle boundaries as well as $\Sigma 9$ CSL boundaries were found susceptible to the initiation of intergranular corrosion in conditions where quantities not exceeding a few equivalent monolayers of material were dissolving. For the $\Sigma 3$ CSL boundaries the behavior was found dependent on the deviation angle of the GB plane from the exact {111} orientation with a transition from resistance to susceptibility occurring between 1° and 1.7° of deviation.

ECSTM was also applied on microcrystalline copper to investigate the local passivation properties at the surface termination of grains boundaries and their relation with the GB type [24,25]. The passive film formed in the Cu(I) passivation region was found to be thicker at grain boundaries than on grains and a thicker Cu(I) passive film was observed at GB edges assigned to random grain boundaries than at GB edges assigned to coherent twins [25]. No metal was observed to be preferentially consumed at GB edges by transient dissolution during Cu(I) passivation. However, the relationship between local passivation behavior and GB structure could not be investigated in further detail because of the lack of detailed characterization of the GB crystallographic parameters at the local sites where passivation could be analyzed at the nanoscale by ECSTM.

Here, we report further insight into the comprehensive knowledge of the GB structure-dependent passivation behavior of metals. The study was performed on microcrystalline copper passivated in the anodic oxidation range where a single $Cu(I)_2O$ oxide film is formed, as shown by *ex situ* [27-32] and *in situ* [33-41] surface analysis. The innovative approach combining ECSTM and EBSD analysis of the same local microstructural area including various types of grain boundaries was applied. To our knowledge, this study is the first combining *in situ* characterization of the local passivation-induced topographic alterations of GB edges with local microstructural characterization of the same GB sites.

## 2. Experimental

Starting from high-purity cast electrolytic tough pitch (ETP-) copper, the microcrystalline samples were produced by cycles of cryogenic rolling and recrystallization annealing [19,22-26,42,43]. Annealing was performed at 200 °C for 1 min in order to limit the grain size and to preserve a GB density suitable for the STM field of view [22,23]. EBSD revealed a nearly random texture of the grains with a $\sum 3$ CSL boundaries length fraction of 66%. The remaining 34% corresponded to random boundaries for the most part [22,23]. Surface preparation was performed by mechanical polishing with diamond spray down to a final grade of 0.25 μm. The cold work layer was removed by electrochemical polishing in 66% orthophosphoric acid at 3 V for 15 s versus a copper counter electrode.

An Agilent Technologies 4500 system comprising PicoSPM base, STM S scanner, PicoScan 2100 controller, PicoStat bi-potentiostat and Picoscan software was used for ECSTM analysis [44-48]. The ECSTM cell exposed a working electrode area of 0.16 cm$^2$ and had two Pt wires serving as counter electrode and pseudo reference electrode, the latter being calibrated *versus* the standard hydrogen electrode (V/SHE). Tips were produced from tungsten rods of 0.25 mm diameter, electrochemically etched and coated with Apiezon wax for electrochemical insulation in the ECSTM cell. A 0.1 M NaOH(aq) aqueous solution was used as electrolyte. It was prepared from ultrapure NaOH and Millipore water (resistivity > 18 MΩ cm).

After exposure to the electrolyte at open circuit potential (-0.03 V/SHE), the samples were submitted to a voltammetry treatment in order to remove the air-formed native oxide and produce an oxide-free metallic state to start with. The potential was cycled at a sweep rate of 0.02 V s$^{-1}$ down to -1.10 V/SHE, at the onset of hydrogen evolution, and backward to the value of -0.60 V/SHE. This treatment was repeated until the cyclic voltammograms (CVs) showed no cathodic peak associated with the reduction of Cu(I) to Cu(0). Two cycles were enough to fully reduce the air-formed native oxide.

The microcrystalline copper surface was then imaged in the metallic state at -0.60 V/SHE in order to include a relevant microstructure in the STM field of view and to localize GB edges of interest. Afterwards, passivation was forced by sweeping (0.005 V s$^{-1}$) the potential up to -0.10 V/SHE in the Cu(I) passivation range. During the sweep, the tip was kept engaged but did not scan the surface so as to keep the very same local area of interest in the STM field of view. Images of the surface in the Cu(I) passive state were then taken at -0.10 V/SHE. Next, the passive oxide film was reduced by cycling the potential (0.02 V/s) down to -1.10 V/SHE and upward to -0.6 V/SHE. For this treatment also, the tip was kept engaged without scanning the surface so as to keep the measurement localized in the very same local area of interest. Images of the surface in the reduced metallic state were then taken at -0.60 V/SHE. All ECSTM images of the surface in the initial metallic, passive and reduced states were acquired in the constant current mode. No filtering was used and the recorded images were processed with the Gwyddion software [49]. In addition to the potential damages to the tip that are inherent to STM measurements, tip alterations could also be induced by the passivation/reduction electrochemical treatments of the surface and were observed in some of the experiments thus limiting the success of obtaining complete series of images of the same local area of interest of the surface in the metallic, passivated and reduced surface state. However, comparative analysis of the data for series of images successfully acquired showed reproducible behavior of the variations of topography measured in the grain boundary regions.

EBSD analysis of the microstructural area of interest analyzed by ECSTM required very precise repositioning of the analyzed area. The repositioning was achieved thanks to micro marks made on the sample surface using the tip as a nano indenter after recording the last STM image. The applied procedure is described in detail elsewhere [26]. A digital scanning electron microscope Ultra55 from ZEISS (FEG-SEM) located at LISE (Laboratoire Interfaces et Systèmes Electrochimiques, CNRS –Université Pierre et Marie Curie) was used for EBSD analysis. The micro marks on the sample surface were first located with SE (secondary electron) images obtained at an accelerating voltage of 20 kV, a nominal beam size of 10 nm and a measured beam current of 1.5 nA. EBSD analysis was then performed on areas of about $30 \times 30$ µm$^2$ including the micro marks. A minimum step size of 0.1 µm was used for acquisition in order to optimize the post-processing for the crystallographic characterization of the grain boundaries, including at the local sites where line profile analysis has been performed by STM. The OIM® software was used for processing the EBSD data.

## 3. Results and discussion

*3.1. Macroscopic electrochemical behavior*

Figure 1 shows a CV characterizing the macroscopic electrochemical behavior of microcrystalline copper in 0.1 M NaOH(aq). It was recorded in the ECSTM cell after reduction of the air-formed native oxide between the onset of hydrogen evolution and the Cu(I) passive range. The anodic AI peak is characteristic for the formation of a Cu(I) oxide layer and the cathodic CI peak for its decomposition to Cu(0) metallic copper [27-41].

Analysis of the charge density transfer was performed for the anodic and reverse cathodic reactions by integrating the current density measured above and below the drawn baseline, respectively. The measured charge densities for the anodic and cathodic reactions were $q_{AI} = 1870\pm10\ \mu C$ cm$^{-2}$ and $q_{CI} = 1710\pm10\ \mu C$ cm$^{-2}$, respectively, showing a partially reversible behavior assigned to the irreversible reaction of copper dissolution occurring during anodic polarization. The equivalent thickness of irreversibly dissolved copper ($\delta_{Cu(IR)}$) can be calculated using Eq. (1) where $q$ is the electric charge density associated to the reaction ($q = q_{AI} - q_{CI}$), $V_M$ the molar volume of the reacting material, $z$ the number of exchanged electrons, and $F$ the Faraday's constant.

$$\delta = \frac{qV_m}{zF} \quad\quad\quad \text{Eq. (1)}$$

One obtains a value of 0.12±0.01 nm of copper ($V_M^{Cu}$ = 7.1 cm³ mol⁻¹, $z$ = 1) being irreversibly dissolved during Cu(I) passivation. This low value amounts to about 0.6 equivalent monolayer (ML) of copper (one (111)-oriented ML of copper is 0.208 nm thick from the bulk *fcc* structure). Using Eq. (1), the equivalent thickness of the passive oxide film ($\delta_{Pass}$) can be calculated from the electric charge density of the reduction peak ($q = q_{CI}$). Assuming the formation of a Cu(I)₂O oxide film ($V_M^{Cu_2O}$ = 23.9 cm³/mol, $z$ = 2 for the reduction of 2 Cu(I)), the value of 2.12±0.01 nm is obtained is in agreement with previous data reported for the Cu(I) passive oxide film formed by cyclic voltammetry [25,27,28,37].

*3.2. Passivation at grain boundaries*

A local area of copper where the microstructure was first imaged *in situ* by ECSTM at the topmost surface and then mapped by EBSD after sample transfer and repositioning of the analysis is shown in Figure 2. The EBSD maps show the micro mark made with the STM tip and used to localize the STM field of view (marked by squares). The EBSD IPF map (Figure 2(b)) shows that the STM field of view includes nine grains labelled G1 to G9. They have different crystallographic orientations except for the smaller grains G5, G6 and G8. The presence of three sub-grains labelled SGA to SGC is also evidenced in grain G2. The SGA and SGB sub-grains have the same crystallographic orientation, different from that of the matrix grain G2. The SGC sub-grain is also differently oriented from the matrix grain G2 and has the same orientation as grain G3.

In the ECSTM image (Figure 2(a)), the surface level for the sub-grains SGA to SGC is systematically measured to be lower than the one of the matrix grain G2, like previously observed [26]. This suggests faster pre-etching of the sub-grains during surface preparation by electrochemical polishing. It is also observed that the grain G3, of same orientation as the sub-grain SGC, has a lower topographic surface level than the surrounding grains G1, G2, G8, G9 but not G4, which is also indicative of faster pre-etching during surface preparation.

The EBSD IQ map (Figure 2(c)) enables us to identify the type of grain boundaries included in the STM field of view. According to this IQ mapping analysis, the boundaries at the G1/G2, G1/G3, G1/G8, G5/G6, G3/G4, G4/G7 and G2/G7 interfaces would be of random type while all others would be grain boundaries of CSL type. Among the CSL boundaries, the boundary at the G8/G9 interface would be $\Sigma 9$ and those at the G2/G3, G4/G5, G4/G6 and G4/G8 interfaces would be $\Sigma 3$ boundaries. All boundaries delimiting the three sub-grains SGA to SGC from the matrix grain G2 would also be $\Sigma 3$ boundaries.

In the ECSTM image of the topmost surface (Figure 2(a)), the grain boundary regions can be identified by the topography variations between the adjacent grains. Like previously reported [22,24-26], line profile analysis revealed that the GB regions could be characterized by the presence of a topographic gap at the interface between the adjacent grains. The sites labelled from 1 to 16 correspond to the local sites where the depth of the topographic gaps, and thus of the GB regions, could be measured in the initial metallic state, in the passive state and in the reduced metallic state. It should be noted that for several boundaries, the line profile measured between the adjacent grains did not reveal a gap sufficiently deep to ensure a reliable measurement of the depth of the GB region and its variation after electrochemical treatment. Such shallow gaps were observed for the random boundaries at the G1/G2, G1/G3, G1/G8, G5/G6, G3/G4, G4/G7 and G2/G7 interfaces and for the CSL boundaries at the G2/G3 and G4/G5 interfaces. Their observation can be assigned to the large difference of topographic level between the adjacent grains masking, at least partially, the topography of the GB edge resolved by the STM tip. Insufficient preferential pre-etching during the electrochemical polishing pre-treatment could also explain their presence.

Like previously reported [24-26], the depth of the GB region was measured from line profiles obtained by averaging 30 adjacent line scans drawn across the GB edge, covering a local distance of ~170 nm along the boundary. The results obtained for the surface in the initial metallic state, in the passive state and in the reduced metallic state are presented in Figure 3. A first examination reveals 3 types of behaviors according to the variation of the depth of the GB region after passivation (Table 1): an increased depth observed in sites 4, 12, 14-16, a decreased depth observed in sites 1-3, 6, 9-11, or an unchanged depth observed in sites 5, 7, 8, 13. An increase of the depth of the GB region can be assigned to the formation of a passive film locally thinner than on the adjacent grains. In the grain boundaries showing this behaviour, passivation would be more efficient since locally consuming less copper than the adjacent grains to form the passive oxide. Inversely, a decrease of the depth of the GB region can be assigned to the formation of a passive film locally thicker and thus to a less efficient passivation consuming more copper than on the adjacent grains by anodic oxidation. The absence of variation of the depth of the GB region in the passive state implies that the passive film is locally as thick as on the adjacent grains and thus that passivation is equally efficient. In nearly all cases the depth measured after reduction of the passive film is identical, within the uncertainty of the measurement, to that measured prior to passivation, indicating a reversible reaction at the GB edges.

Let us now discuss the results for the different grain boundaries observed in Figure 2 and start with the grain boundaries delimiting the sub-grains SGA and SGB from grain G2. These boundaries are all of Σ3 CSL type according to the EBSD IQ map. We observe a similar behaviour characterized by an increase of the GB depth in the passive state at sites 12 and 14 (delimiting sub-grain SGA) and at sites 15 and 16 (delimiting sub-grain SGB). However, at site 13, corresponding to another grain boundary delimiting the sub-grain SGA, no significant variation is observed. In all cases, the GB depth variation is reversible after reduction of the passive film.

The grain boundary measured at sites 2 to 6 along the G4/G8 interface is also of Σ3 CSL type according to the EBSD IQ map. It shows a varying behaviour since, at sites 2, 3 and 6, a significant decrease of the GB depth is measured after passivation whereas a very small increase or no variation are measured at sites 4 and 5, respectively. The depth variations measured along this boundary are either reversible (site 2 and 5) or irreversible (sites 3, 4 and 6) after reduction of the passive film. The Σ3 CSL boundary measured at sites 7 to 11 along the G4/G6 interface also shows a varying behaviour since a significant decrease of the GB depth is measured at sites 9 to 11 but not at sites 7 and 8. After reduction of the passive film, the observed depth variations are reversible. For the boundary measured at site 1 across the G8/G9 interface and of Σ9 CSL type according to the EBSD IQ map, the depth decrease is the largest observed after passivation and the change is reversible after reduction of the passive film.

Thus, for grain boundaries that are classified of the same Σ3 CSL type by the EBSD IQ mapping analysis, the efficiency of the passivation varies as revealed by ECSTM analysis of the increased, decreased or unchanged depths of the GB regions. Clearly these results suggest that although of the same Σ CSL type, the local GB structure may vary between different boundaries and also along the same boundary, modifying the local reactivity towards passivation similarly to what was observed in the initial stages of active dissolution [26].

*3.3. Effect of GB local structure*

In order to better rationalize the effect of the GB structure, the EBSD analysis was refined. At each local site where line profile analysis of the ECSTM data was performed, the misorientation angle $\theta$ and the deviation angle $\varphi$ of the boundary with respect to the ideal coherent GB geometry were extracted. The results are compiled in Table 1.

At sites 1 and 2, the local EBSD analysis did not yield values of the misorientation angle consistent with a well-defined Σ CSL type, meaning that the GB local structure is of random type. This local random type of the structure of the boundary can thus be used to explain the locally lower efficiency for passivation and thus higher reactivity of the boundary. This result is consistent with the higher reactivity (i.e. susceptibility) of random type boundaries observed in the initial stages of active dissolution [26] and in later stages of sub-surface penetrating attack [2,6,8,11-14,16]. Thus, random type high angle boundaries are susceptible to intergranular attack in the absence of the passive film and are not efficiently self-protected by passivation.

At sites 3 to 16, all boundaries can be classified as Σ3 CSL type in agreement with the EBSD IQ mapping analysis. The Brandon [50] and Palumbo-Aust criteria [51] define ranges of tolerance of the misorientation angle outside of which the boundary can no longer be considered as a specific Σ CSL. For the Σ3 misorientation, these ranges are 60±8.67° and 60±6°, respectively. For the sites 3 to 16, these criteria are respected and the GB structure is locally a $\Sigma 3$ CSL one. Thus, the variations of the passivation behaviors measured at these sites are hereafter discussed based on the deviation angle of the GB plane, i.e. in terms of coherent vs. incoherent local structure of the boundary.

Along the G4/G8 interface (sites 2 to 6), the largest deviation ($\varphi = 1°$) with respect to the perfect $\Sigma 3$ CSL geometry ($\varphi = 0°$) is observed at site 6, which would explain the higher reactivity of the site and thus its lower efficiency to passivate and the formation of a locally thicker passive film. At sites 3 to 5, the deviation with respect to the perfect geometry is lower ($\varphi = 0.4$-$0.5°$), which would imply a lower reactivity of the sites and thus their better efficiency to passivate as indicated by the observed variability of the depth variations of the GB region.

Along the G4/G6 interface (sites 7 to 11), the local deviation of the boundary with respect to the perfect geometry is $0.5°$ at the most. This also implies a lower reactivity and a better efficiency to passivate than at site 6, which is indeed observed at sites 7 and 8 ($\varphi = 0.3°$) where no variation of the depth of the GB region is observed. At sites 9 to 11 for which the local deviation is higher ($\varphi = 0.4$-$0.5°$), a higher reactivity and thus lower efficiency to passivate is measured as indicated by the decrease of the depth of the GB region. One can notice in Figure 2 that, in this region, the GB edge presents a kinked morphology indicative of the accumulation of misfit dislocations in the GB plane, and contributing to its deviation from the perfect geometry.

At the SGA/G2 (site 12 to 16) interfaces, the local deviation of the boundary with respect to the perfect geometry is $0.3°$ at the most. Note that such low values of the deviation angle $\varphi$ are indicative of nearly perfect coherent twins, which is also suggested by the presence of parallel GB edges at the surface (Figure 2). The lower deviated local structure of these coherent twins implies a lower reactivity and increased efficiency to passivate which is confirmed by the observed variations of the GB depth measured at sites 12 to 16, where the passive film was measured to be thinner than on the adjacent grains in most cases.

In our previous study of active dissolution of microcrystalline copper [26], the transition from resistance to susceptibility to the initiation of the intergranular attack of Σ3 CSL boundaries was observed for a deviation angle φ of 1.0-1.7°. This value was concluded to correspond to the transition from resistive coherent twins to susceptible incoherent twins in the absence of formation of a passive film. The transition was explained by the presence of misfit dislocations (i.e. steps) accommodating the deviation of the boundary plane from the exact CSL plane and by the increased density of step regions introduced in the boundary plane with increasing deviation [26]. The present data obtained for copper passivation by formation of a Cu(I) oxide film show a transition from more efficient to less efficient passivation for a deviation angle φ of 0.4-0.5°. This implies that the effect of the presence of an increasing density misfit dislocations (i.e. steps) in the boundary plane is more detrimental for the efficiency to passivate than for the resistance to dissolve.

## 4. Conclusions

The relationship between grain boundary type and structure and passivation efficiency was studied for the first time by combining ECSTM and EBSD analysis of the same microstructural area. The study was performed on microcrystalline copper in 0.1 M NaOH(aq) aqueous solution. The *in situ* ECSTM data show that passivation by anodic oxidation in the Cu(I) range is characterized in most cases by a decrease of the depth of GB edge region due to the formation of a passive film locally thicker than on the adjacent grains and thus to anodic oxidation being locally less efficient since consuming more copper at the grain boundaries. Cu(I) passivation is less efficient at the edges of high angle random boundaries as compared Σ3 CSL boundaries.

Coupling ECSTM and EBSD enabled us to address the effect of the GB fine structure on the passivation properties. For Σ3 CSL boundaries, a transition from more efficient to less efficient passivation was observed for a deviation angle φ of 0.4-0.5° of the {111} GB plane with respect to the perfect geometry of an ideal coherent twin boundary. For lower values of the deviation angle, the GB edges would be less reactive and passivate more efficiently than adjacent grains whereas for higher values the local behaviour can be either identical to that of the adjacent grains or similar to that of random boundaries. This transition would result from the presence of misfit dislocations (i.e. steps) accommodating the deviation of the boundary plane from the exact CSL plane and from the effect of an increased density of steps introduced in the boundary plane with increasing deviation on the local anodic oxidation properties.

## Acknowledgments

This project has received funding from the European Research Council (ERC) under the European Union's Horizon 2020 research and innovation programme (ERC Advanced Grant No 741123, Corrosion Initiation Mechanisms at the Nanometric and Atomic Scale, CIMNAS).

# Table

Table 1 Grain boundary crystallographic characteristics as measured by EBSD at the sites marked in Figure 2 and variation of the depth of the GB region and efficiency of Cu(I) passivation behavior with respect to adjacent grains as measured by ECSTM.

| Site | GB type | GB misorientation angle $\theta$ (°) | Deviation angle $\varphi$ (°) | Variation of GB depth | Passivation efficiency |
|---|---|---|---|---|---|
| 1 | R | ----- | ----- | ↓ | − |
| 2 | R | ----- | ----- | ↓ | − |
| 3 | $\Sigma 3$ | 59.8 | 0.4 | ↓ | − |
| 4 | $\Sigma 3$ | 59.7 | 0.4 | ↑ | + |
| 5 | $\Sigma 3$ | 59.5 | 0.5 | ≈ | ≈ |
| 6 | $\Sigma 3$ | 59.6 | 1 | ↓ | − |
| 7 | $\Sigma 3$ | 60 | 0.3 | ≈ | ≈ |
| 8 | $\Sigma 3$ | 59.9 | 0.3 | ≈ | ≈ |
| 9 | $\Sigma 3$ | 59.4 | 0.5 | ↓ | − |
| 10 | $\Sigma 3$ | 59.6 | 0.4 | ↓ | − |
| 11 | $\Sigma 3$ | 59.8 | 0.4 | ↓ | − |
| 12 | $\Sigma 3$ | 60 | 0.3 | ↑ | + |
| 13 | $\Sigma 3$ | 60 | 0.1 | ≈ | ≈ |
| 14 | $\Sigma 3$ | 60 | 0.2 | ↑ | + |
| 15 | $\Sigma 3$ | 60 | 0.3 | ↑ | + |
| 16 | $\Sigma 3$ | 60 | 0.3 | ↑ | + |

**Figure captions**

Figure 1 Cyclic voltammogram recorded in the ECSTM cell for microcrystalline copper in 0.1 M NaOH(aq), scan rate = 20 mV/s.

Figure 2 Local microstructure of polycrystalline copper imaged *in situ* by ECSTM and then mapped by EBSD after repositioning in the same area: (a) Topographic ECSTM image of the metallic state at E = -0.6 V/SHE in 0.1 M NaOH(aq) (Z range $\Delta Z$ = 15 nm, tip potential $E_{tip}$ = -0.9 V/SHE, tunneling current $I_t$ = 1.5 nA); (b) EBSD inverse pole figure (IPF) map; (c) EBSD image quality (IQ) map. In (a) and (b), grains are labelled G1 to G9 and sub-grains SGA to SGC. The local GB sites selected for depth profile analysis are indexed 1 to 16. In (b) and (c), the squares mark the field of view analyzed by STM.

Figure 3 Bar graph of the depth measured across the grain boundaries at the sites # 1 to 16 in Figure 2(a) in the initial metallic state at E = -0.6 V/SHE, in the passive state at E = -0.1 V/SHE and in the final metallic state at E = -0.6 V/SHE after reduction of the passive film. The symbols ↓, ↑ and ≈ denote a lower, higher or equivalent local passivation efficiency, respectively, as deduced from the variations of the GB depth after passivation.

**Figure 1**

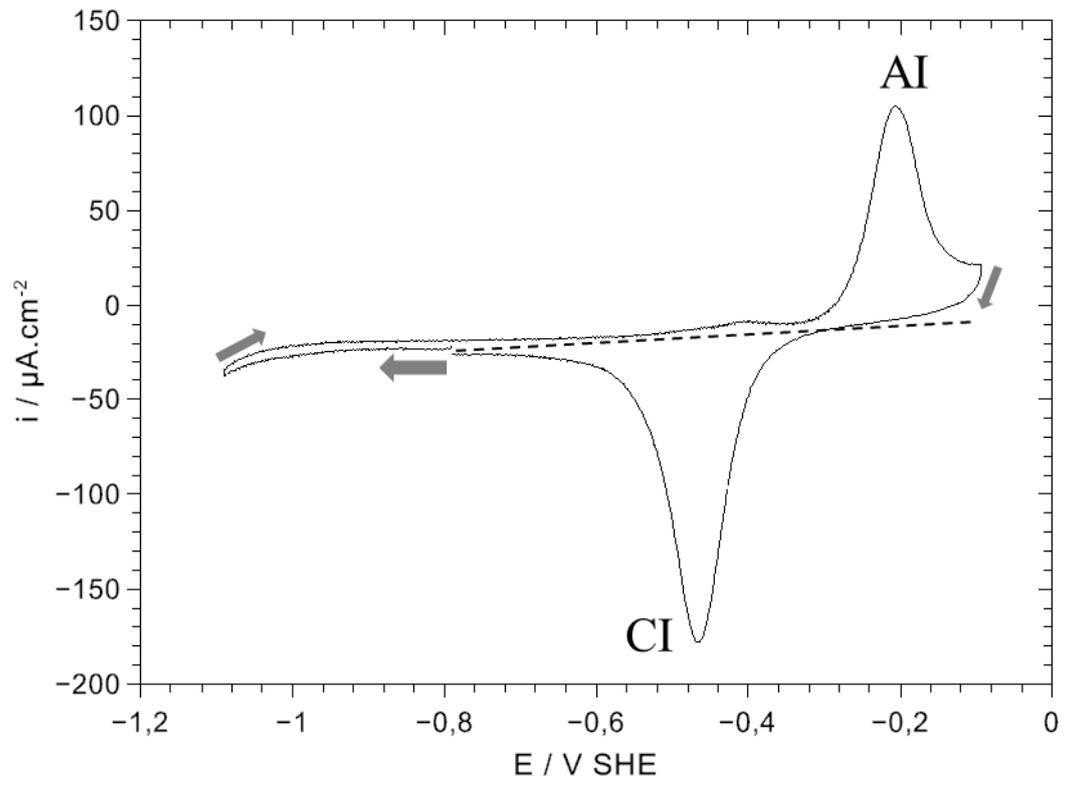

**Figure 2**

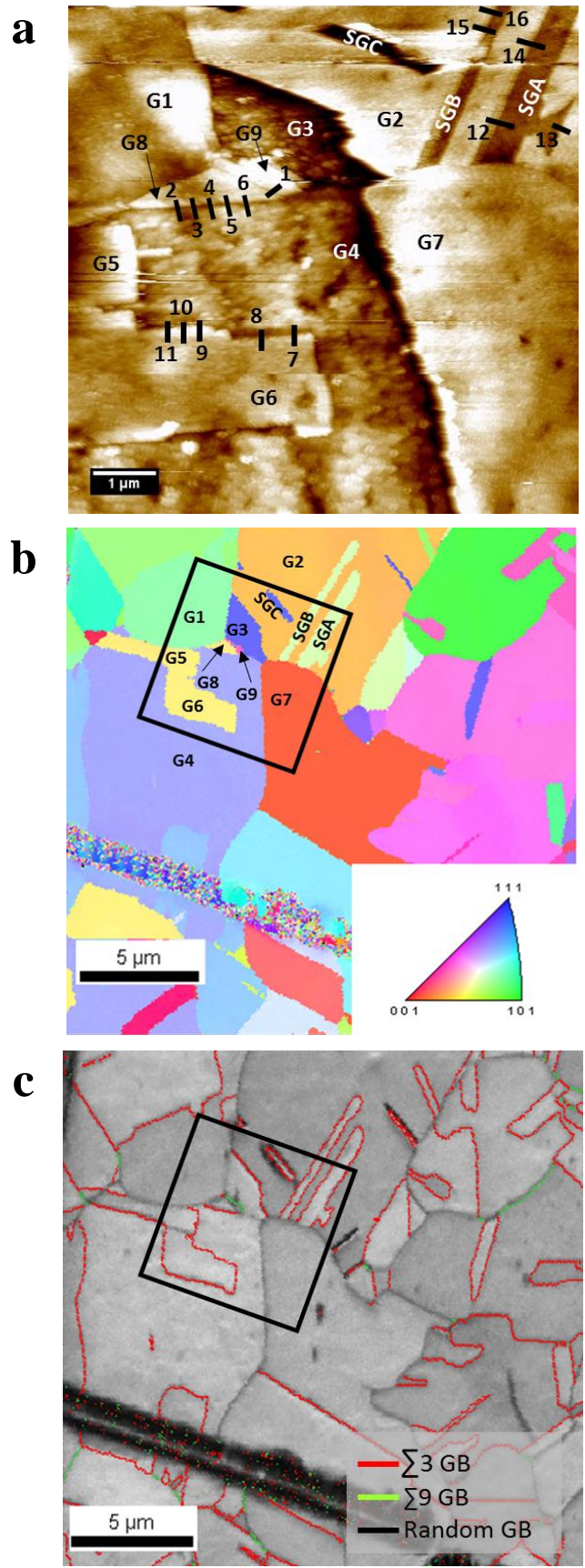

**Figure 3**

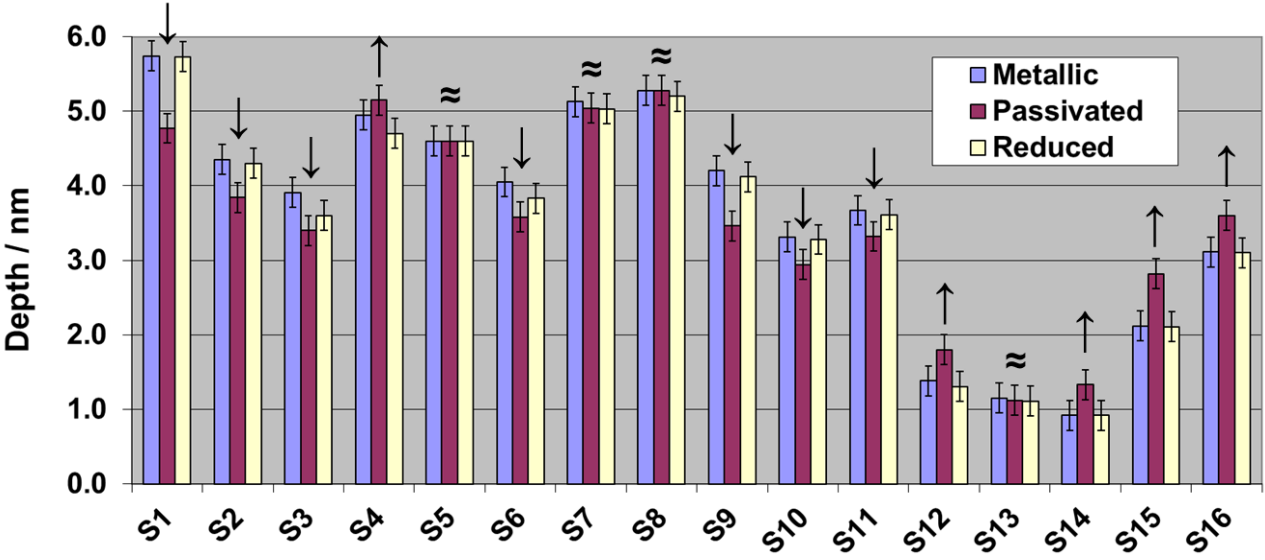